\documentclass{emulateapj}
\usepackage{dsfont}

\newcommand{\myemail}{e-Mail: havenhaus@gmail.com}

\slugcomment{Accepted for publication in Astrophysical Journal on May 25, 2017}

\shorttitle{Exploring dust around HD142527 down to 0.025\arcsec / 4au using SPHERE/ZIMPOL}
\shortauthors{Avenhaus et al.}

\begin{document}

\title{Exploring dust around HD142527 down to 0.025\arcsec / 4au using SPHERE/ZIMPOL}

\author{
H.~Avenhaus$^{1, 2, 3, 4}$, 
S.P.~Quanz$^{2}$, 
H.M.~Schmid$^{2}$, 
C.~Dominik$^{5}$, 
T.~Stolker$^{5}$, 
C. Ginski$^{6}$, 
J.~de~Boer$^{6}$, 
J.~Szul\'agyi$^{2}$, 
A.~Garufi$^{7}$, 
A.~Zurlo$^{4,8}$, 
J.~Hagelberg$^{9}$, 
M.~Benisty$^{9}$, 
T.~Henning$^{10}$, 
F.~M\'enard$^{9}$, 
M.~R.~Meyer$^{11}$,
A.~Baruffolo$^{12}$, 
A.~Bazzon$^{2}$, 
J.L.~Beuzit$^{13,14}$, 
A.~Costille$^{15}$, 
K.~Dohlen$^{15}$, 
J.H.~Girard$^{16}$, 
D.~Gisler$^{17}$, 
M.~Kasper$^{18}$, 
D.~Mouillet$^{13,14}$, 
J.~Pragt$^{19}$, 
R.~Roelfsema$^{19}$, 
B.~Salasnich$^{12}$, and
J.-F.~Sauvage$^{15}$}

\email{\myemail}
\altaffiltext{1}{Based on observations collected at the European Organisation for Astronomical Research in the Southern Hemisphere, Chile, as part of the SPHERE GTO observations.}
\altaffiltext{2}{ETH Z¬urich, Institute for Astronomy, Wolfgang-Pauli-Strasse 27, CH-8093, Z¬urich, Switzerland}
\altaffiltext{3}{Departamento de Astronom\'ia, Universidad de Chile, Casilla 36-D, Santiago, Chile}
\altaffiltext{4}{Millennium Nucleus "Protoplanetary Disk", Departamento de Astronom\'ia, Universidad de Chile, Casilla 36-D, Santiago, Chile}
\altaffiltext{5}{Anton Pannekoek Institute for Astronomy, University of Amsterdam, Science Park 904, 1098 XH Amsterdam, The Netherlands}
\altaffiltext{6}{Leiden Observatory, Leiden University, P.O. Box 9513, 2300 RA Leiden, The Netherlands}
\altaffiltext{7}{Universidad Aut\'onoma de Madrid, Dpto. F\'isica Te\'orica, M\'odulo 15, Facultad de Ciencias, Campus de Cantoblanco, E-28049 Madrid, Spain}
\altaffiltext{8}{N\'ucleo de Astronom\'ia, Facultad de Ingenier'a, Universidad Diego Portales, Av. Ejercito 441, Santiago, Chile}
\altaffiltext{9}{Univ. Grenoble Alpes, CNRS, IPAG, F-38000 Grenoble, France}
\altaffiltext{10}{Max-Planck-Institut fur Astronomie, K\"onigstuhl 17, D-69117 Heidelberg, Germany}
\altaffiltext{11}{Department of Astronomy, University of Michigan, 1085 S. University, Ann Arbor, MI 48109, USA}
\altaffiltext{12}{INAF Ð Osservatorio Astronomico di Padova, Vicolo dell'Osservatorio 5, 35122 Padova, Italy}
\altaffiltext{13}{Universit\'{e} Grenoble Alpes, IPAG, 38000 Grenoble, France}
\altaffiltext{14}{CNRS, IPAG, 38000 Grenoble, France}
\altaffiltext{15}{Aix Marseille Universit\'{e}, CNRS, LAM, UMR 7326, 13388, Marseille, France}
\altaffiltext{16}{European Southern Observatory, Alonso de Cordova 3107, Casilla 19001 Vitacura, Santiago 19, Chile}
\altaffiltext{17}{Kiepenheuer-Institut f\"{u}r Sonnenphysik, Schneckstr. 6, D-79104 Freiburg, Germany}
\altaffiltext{18}{European Southern Observatory, Karl Schwarzschild St, 2, 85748 Garching, Germany}
\altaffiltext{19}{NOVA Optical Infrared Group, Oude Hoogeveensedijk 4, 7991 PD Dwingeloo, The Netherlands}

\begin{abstract}

We have observed the protoplanetary disk of the well-known young Herbig star HD 142527 using ZIMPOL Polarimetric Differential Imaging with the VBB (Very Broad Band, $\sim$ 600-900nm) filter. We obtained two datasets in May 2015 and March 2016. Our data allow us to explore dust scattering around the star down to a radius of $\sim$0.025$\arcsec$ ($\sim$ 4au). The well-known outer disk is clearly detected, at higher resolution than before, and shows previously unknown sub-structures, including spirals going inwards into the cavity. Close to the star, dust scattering is detected at high signal-to-noise ratio, but it is unclear whether the signal represents the inner disk, which has been linked to the two prominent local minima in the scattering of the outer disk, interpreted as shadows. An interpretation of an inclined inner disk combined with a dust halo is compatible with both our and previous observations, but other arrangements of the dust cannot be ruled out. Dust scattering is also present within the large gap between $\sim$30 and $\sim$140au. The comparison of the two datasets suggests rapid evolution of the inner regions of the disk, potentially driven by the interaction with the {close-in M-dwarf companion}, around which no polarimetric signal is detected.

\end{abstract}

\objectname{HD142527} 

\section{Introduction}

{Planet formation cannot be understood from theory and first principles alone. Besides looking at the specific case of the solar system , the most promising way to achieve better comprehension is to study the environments planets form in - before, during, and after their formation. }Protoplanetary and specifically transition disks, {i.e. disks in which the inner region has already undergone some clearing \citep[for a recent review, see][]{espaillat2014}}, are interesting targets when trying to examine and understand the intricate processes that occur in the circumstellar environment in its early phases.
We know that planets form during these early phases in circumstellar disks. It is suspected that the transition disk phase is directly related to the formation of planets, because often, planets are {a possible explanation} for the features seen in the disk \citep{andrews2011, espaillat2012}. 

In a few cases, the connection between the disk and the (forming) planets has already been made. In HD~169142, a companion candidate was detected just inside the inner rim of the outer disk \citep{reggiani2014, biller2014}, and an outer clump was seen at mm wavelengths \citep{osorio2014}. In LkCa15, two companion candidates have been detected by means of non-redundant masking and/or direct H$\alpha$ imaging \citep{kraus2012, sallum2015}. In HD~100546, a companion candidate 
potentially in its accretion phase was detected at $\sim$53$\pm$2 au, and a second one inside the cleared region is suspected based on the orbital notion seen in CO ro-vibrational line spectra and the shape of the inner rim of the outer disk \citep{quanz2013a, quanz2015, mulders2013, brittain2013, brittain2014, montesinos2015}. All these disks are transition disks, and in all cases except for the outer HD~100546 companion the physical connection between the companion and the disk seems reasonably clear, strengthening the interpretation of transition disks as an important evolutionary phase related to the birth of at least certain types of planetary systems. Structures in protoplanetary disks are generally often interpreted in terms of planets or companions, which can induce gaps or spiral arms \citep[e.g.][]{casassus2015a, perez2015, dong2015, dong2016}. However, it is important to keep in mind that structures in disks can also be produced by processes not necessarily involving planets or planet formation, such as grain growth, photo-evaporation, magneto-rotational instabilities, shadow-driven spirals, or vortices \citep[e.g.][]{chiang2007, owen2011, ataiee2013, montesinos2016, baruteau2016, ragusa2016}.

Many transition disk systems are still accreting \citep{sicilia2010, fairlamb2015}, though perhaps at lower than expected levels \citep{najita2015}. This means that they must have an inner accretion disk, and given the typical clearing times of these inner disks, it also means that in most cases, material must be able to cross the gap to feed the accretion. In several cases, these inner disks have been detected directly with near-IR interferometry / imaging in the mid-IR or inferred from their SEDs \citep[e.g.][Pineda et al. in prep, Szulagyi et al. in prep.]{fedele2008, benisty2010, tatulli2011, mulders2013, schegerer2013, olofsson2013, maaskant2013, menu2014, panic2014, matter2016}. In the case of LkCa15, an inner disk has been directly detected in scattered light \citep{thalmann2016, oh2016, thalmann2015}, but this is generally a challenging undertaking because of the proximity to the star and the contrasts needed. The dust emission is often too faint to be seen in (sub-)mm even with ALMA, though an unresolved inner disk has been inferred for the nearest protoplanetary transition disk, TW Hya, {which could be responsible for a rotating azimuthal asymmetry seen in scattered light  \citep{andrews2016, vanboekel2017, debes2017}}.

\subsection{The HD142527 system}

HD~142527 is {an example of a Herbig star surrounded by an optically thick protoplanetary disk and well-studied }due to its proximity \citep[$156^{+7}_{-6}$ pc,][]{gaia2016}, {the brightness of its disk} \citep[F$_{\rm IR}$/F$_{\rm star}$ = 0.92,][]{verhoeff2011} and the size of its gap ($\sim$140 au given the new Gaia distance). {It is only moderately inclined \citep[estimates range from $\sim$20$^\circ$ to $\sim$27$^\circ$,][]{pontoppidan2011, boehler2017}}. These factors together have allowed for high-resolution, high-SNR images of the disk in optical/near-IR scattered light \citep{fukagawa2006, rameau2012, casassus2012, canovas2013, casassus2013, avenhaus2014a, rodigas2014}, the mid-infrared \citep{leinert2004} and at sub-mm wavelengths \citep{casassus2013, casassus2015a, perez2015, muto2015}. The disk has also been studied {with respect to} its sub-mm polarization \citep{kataoka2016}. These studies have revealed the gap to be highly depleted in micron-sized grains, though optically thick CO gas is present. They have also led to the interpretation of the two prominent local minima in scattered light as shadows cast from an inner, tilted disk, in agreement with the ALMA studies of gas close to the star \citep{marino2015, casassus2015b}. More recently, \citet{min2016} have produced an updated model of the inner and outer disk based on Herschel data, focusing on the water ice within the disk. Spirals in the outer disk extend outward from its inner rim and are detected both in sub-mm and scattered light \citep{fukagawa2006, christiaens2014}, though they are displaced w.r.t. each other at the two wavelengths. The inner disk has so far been revealed through mid-IR imaging and SED modeling, but has not been imaged directly in the near-IR at high resolution (sub-0.1$\arcsec$). {The disk still retains significant mass \citep[total disk mass of $\sim$0.1-0.15 M$_\odot$,][]{acke2004a, fukagawa2006} and besides the fact that there is no good agreement on the extinction towards or the exact luminosity of the star, the star is still accreting at a significant rate \citep[$\dot{\rm M} = 6.9\cdot10^{-8} \,{\rm M}_\sun{\rm yr}^{-1}$,][]{lopez2006}, which means that material must be transported in some way from the outer to the inner regions of the disk.}

\begin{figure*}
\centering
\includegraphics[width=0.95\textwidth]{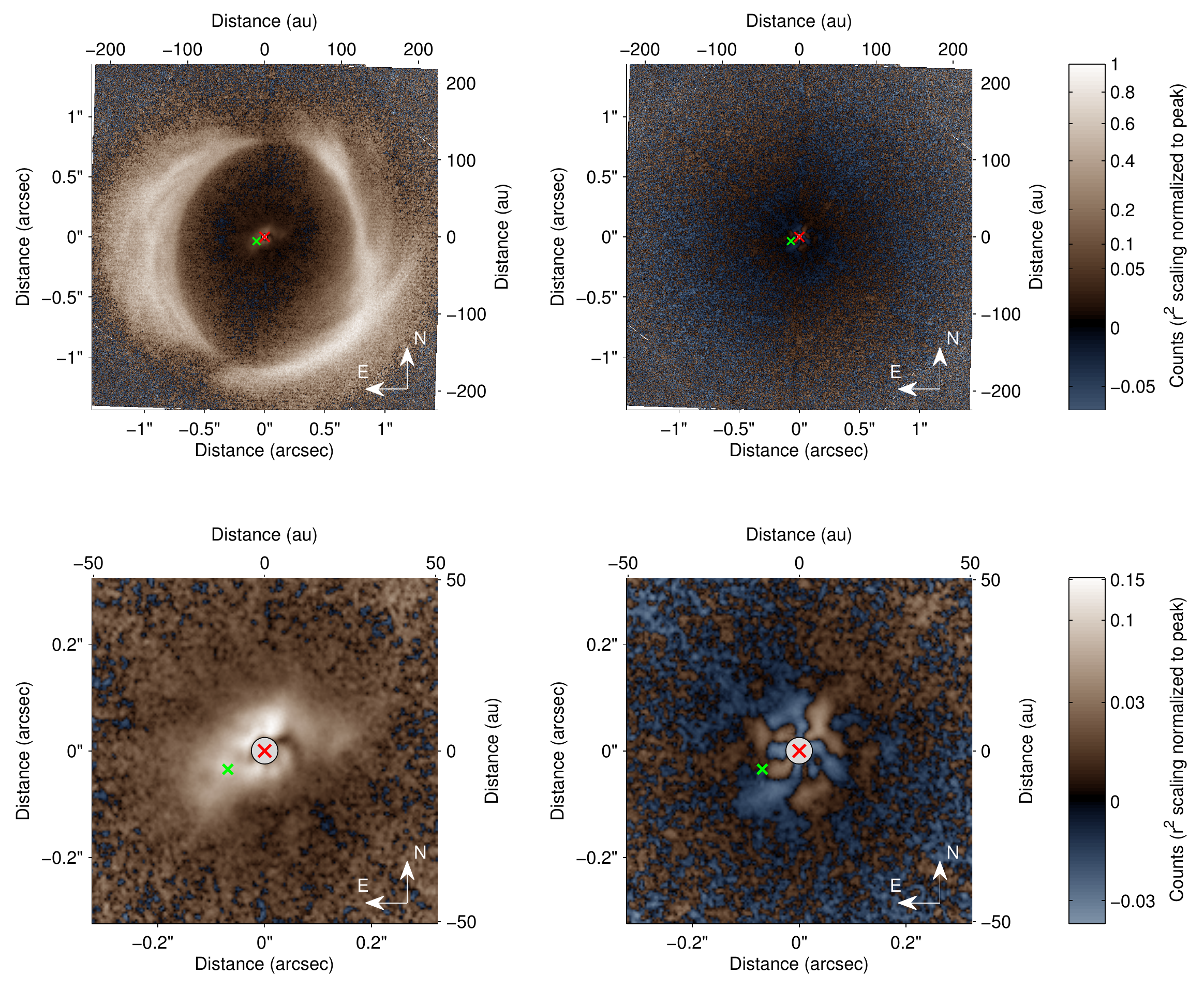}
\vspace{8pt}
\caption{Final $Q_{\phi}$ and $U_{\phi}$ images, shown in the same stretch ({$r^2$ scaling normalized to peak intensity and using a square-root stretch, }2015 and 2016 data combined). {Top two panels: Entire field of view. Bottom two panels: Same data, but zoomed in to the inner regions and enhanced (see color bar)}. Orange hues denote positive, blue hues negative values. The {red} X marks the position of the primary, the small green X the position of the secondary (latest published position of May 12, 2014, \citet{lacour2016}, {separation 77.2 mas}). {Data within our inner working angle of 25 mas has been masked out. Note the square-root stretch, which reduces overall contrast, but better shows faint features of the disk and also enhances the visibility of noise in the $U_{\phi}$ images}.}
\label{figOverview}
\end{figure*}

So far, the inner disk surrounding the star has escaped detections in scattered light. Dust close to the star (up to to $\sim$30 au) is consistent with SED modelling and mid-IR imaging \citep{verhoeff2011}, but using NACO, \citep{avenhaus2014a} reached an inner working angle of $\sim$0.1$\arcsec$ (15 au) without detecting traces of the inner disk. This is interesting in light of the fact that HD~142527 has a still accreting M-dwarf companion, which has been detected with NACO/SAM and subsequently been followed up with NACO, GPI, MagAO, and most recently with SPHERE at close separation (77-90 mas). While it does not seem to orbit in the same plane as the outer disk, its orbit could be in agreement with the orientation of an inner disk casting shadows, though it is still not very well-determined {\citep[][Quanz et al. in prep.]{biller2012, close2014, lacour2016}}. This companion must necessarily be in causal contact and thus shape the dust and gas present within the inner 10-30 au around the star. Recent observations by \citet{rodigas2014} performed with the Gemini Planet Imager (Y band, 0.95-1.14$\mu$m) detect the companion in total intensity and find a point source in polarized light slightly offset from the location of the secondary, suggesting the presence of dust close to the location of the M-dwarf companion, possibly in a circumsecondary disk.

Here, we present new SPHERE/ZIMPOL observations of HD~142527 specifically designed to study the smallest possible separations at the highest possible detail and SNR, being able to detect and resolve circumstellar dust down to an inner working angle (IWA) of $\sim$25 mas (4 au). In Section 2, we describe the data and data reduction procedures. In Section 3, we present results and analysis of these data, which are then discussed in more detail in Section 4. We conclude in Section 5.

\section{Observations and data reduction}\label{observations_section}

\label{Observations}

The observations were performed at the Very Large Telescope on the night of May 2nd, 2015, as part of the SPHERE GTO campaign. A second set of observations was obtained on March 31st, 2016. The ZIMPOL sub-instrument of SPHERE \citep{beuzit2008, thalmann2008} was used in polarimetric differential imaging (PDI) mode, with both instrument arms set up with the Very Broad Band (590-881nm) filter, covering a wide wavelength range from the R to the I band. The read-out mode was set to FastPol with an integration time of 3s per frame, very slightly saturating the PSF core (2s and no saturation for the 2016 data). The data were taken using the P2 polarimetric mode (field stabilized) of ZIMPOL in 2015, and using the P1 (not field stabilized, i.e. the field rotates) and P2 modes for equal amounts of time in 2016 (this was done in order to compare the performance of the P1 and P2 modes for ZIMPOL). The instrument was set up to maximise the flux on the detector while enabling the study of the very smallest separations (down to $\sim$25 mas) and the gap of the disk as deeply as possible. The setup is not optimised for a high-SNR detection of the outer disk.

In order to minimize suspected systematic effects close to the star (see section \ref{secInnerDust}), images were taken in four blocks with different de-rotator angles in 2015 (0, 35, 80, 120 degrees). In 2016, the two polarimetric modes used naturally gave two different field rotations, with the field slightly rotating ($\sim$4 degrees) in the P1 mode. The P1 and P2 mode observations were interleaved, with 3 blocks for each. For each of these blocks, the number of integrations (NDIT) was set to 14 (18 for the 2016 data), with the number of polarimetric cycles (NPOL) set to 6 (5 and 4 for the two last rotations after frame selection, 6 for the 2016 data). Using the QU cycle (full cycle of all four half-wave plate rotations), this adds up to a total on-source integration time of 3528s (3s * 14 (NDIT) * (6+6+5+4 (NPOL)) * 4 (HWP rotations)) in 2015 and 5184s (2s * 18 (NDIT) * 6 (NPOL) * 4 (HWP rotations) * 3 (blocks) * 2 (P1+P2)) in 2016, for a grand total of 8712s (2h 25.2min) of on-source integration time in both epochs combined.

The most critical step in PDI is the centering of the individual frames. ZIMPOL data are special because of the way the detectors work \citep[see][and the SPHERE user manual]{thalmann2008}. The pixels cover an on-sky area of ~7.2mas x 3.6mas each. The stellar position is determined before re-scaling the images by fitting a skewed (i.e. elliptical) 2-dimensional Gaussian to the peak. The data are then re-mapped onto a square grid, accounting for the difference in pixel scale along the x- and y-axis, and corrected for True North. The columns affected by the read-out in FastPol mode have been mapped out manually \citep[see also][]{schmid2012}. Besides these points specific to the ZIMPOL detectors, the data reduction follows the steps described in \citet{avenhaus2014a} with one important difference: Instead of performing the correction for instrumental (or interstellar) polarization for each pair of ordinary and extraordinary beams, the correction is done at the end by subtracting scaled versions of the total intensity I from the Stokes Q and U vectors, minimizing the absolute value of $U_{\phi}$. This method has been used successfully before by the SEEDS team \citep{follette2013} {and is better at suppressing very low surface brightness residuals. We note, however, that any such method that does not rely on separate calibration sources intrinsically assumes the star to be unpolarized. Any intrinsic polarization of the central source will diminish the resultant data quality. However, given the typical polarizations of Herbig stars (fractions of percent to few percent) compared to the polarization induced by dust scattering \citep[10-50$\%$, e.g.][]{avenhaus2014a}, this would be a second-order effect.} A correction for the efficiency in Stokes U vs. Stokes Q is not required for ZIMPOL thanks to the to the better controlled instrumental polarization compared to NACO \citep[no significant crosstalk from Stokes U to Stokes I;][]{bazzon2012}.

The local Stokes vectors, now called $Q_{\phi}$ and $U_{\phi}$ by most authors \citep[e.g.][]{benisty2015} are calculated as:

$$Q_{\phi}=+Q\cos(2\phi)+U\sin(2\phi)$$
$$U_{\phi}=-Q\sin(2\phi)+U\cos(2\phi)$$
$$\phi=\arctan\frac{x-x_0}{y-y_0}+\theta$$

Here, $\theta$ is used to correct for the fine-alignment of the half-wave plate (HWP) rotation and is determined from the data by assuming that $U_{\phi}$ should on average be zero. {We note that in cases of highly inclined optically thick disks, the reality can deviate from this assumption strongly due to multiple scattering \citep{canovas2015}. HD142527 is only moderately inclined \citep{pontoppidan2011}, but the inner disk is suspected to be inclined by about 70 degrees \citep{marino2015}. However, in an optically thick, but symmetric disk of any inclination, the average of $U_{\phi}$ will still be zero for reasons of symmetry. In the case of single-scattering (optically thin disks) and non-aligned grains, the assumption of no signal in $U_{\phi}$ (polarization signal perpendicular to the incident light) holds for any inclination.}

{The disk of HD142527 is neither optically thin nor symmetric. We use the region between 0.2\arcsec and 0.6\arcsec to measure $U_{\phi}$ for correction. This region has very little flux in either $Q_{\phi}$ or $U_{\phi}$, resulting in a good correction for instrumental or interstellar polarization effects. Given the inherent problems with measuring flux in PDI images \citep{avenhaus2014b}, we do not attempt to perform an absolute flux calibration of our images.} 

{We roughly estimate the Strehl ratio of our data by comparing the flux within the first airy minimum to the total flux (measured within 1.5") and dividing this by the expected ratio for a perfect, diffraction-limited system of 0.838. Using this method, we arrive at an estimate of around 34\% for all our datasets. The resolution achieved (as measured by the FWHM) is around 34 mas, again for all our datasets.}

\section{Results and Analysis}\label{results}
\label{sec:results}

In this section, we first discuss the results for the combined data of both epochs, {before investigating} possible differences between both epochs in section \ref{sec:2015vs2016}.

Figure \ref{figOverview} shows the resulting combined (2015+2016) $Q_{\phi}$ and $U_{\phi}$ images obtained in the same color stretch. To first order, $Q_{\phi}$ contains polarimetric signal and noise, while $U_{\phi}$ contains no signal, but noise on the same level. 

As can be seen, the signal in $Q_{\phi}$ is - as expected - much stronger than the signal in $U_{\phi}$, and polarized flux is seen close to the star. However, there remains a significant pattern close to the star in $U_{\phi}$ within the innermost $\sim$200 mas. This also affects the $Q_{\phi}$ data. This kind of noise is seen for other sources as well \citep[e.g. {HD 135344B},][]{stolker2016}. Taking the data in different orientations reduces this problem, but does not eliminate it. {It is worth noting that this pattern noise, which is overlaid over the actual data in both $Q_{\phi}$ and $U_{\phi}$, has both positive and negative components. There is no indication in either this or the HD 135344B dataset that the noise deviates from zero on average, and thus the noise mostly cancels itself out once taking azimuthal averages.}
The dust scattering close to the star remains a clear detection and it is evident in each of our 6 independent datasets individually. Its significance is further emphasized by statistical analysis (see below). Besides the noise pattern, we do not detect any significant astrophysical signal in $U_{\phi}$.

\subsection{Geometrical appearance of the dust}

The $Q_{\phi}$ images reveal the well-known outer disk at high SNR and at higher resolution than previously available NACO data \citep{canovas2013, avenhaus2014a}. They furthermore are able to detect dust scattering close to the star. This structure is elongated in the ESE-WNW (position angle $\sim$120$^\circ$ east of north) direction. However, it does not  resemble a uniform disk of any inclination, {but rather has extensions both on the south-eastern and western / north-western sides (these two lobes are seen in each of the six independent sub-datasets described before). The more prominent extension is seen on the north-western side. The western side is also special because there is a prominent dip in brightness towards the west, very close to the star ($\sim$25-50mas). The dust structure as a whole has no apparent relation to the shadows seen in the outer disk \citep{marino2015}, because an inclined disk explaining those shadows would be elongated in the north-south direction.}

Furthermore, the gap that was first revealed to be largely devoid of dust down to $\sim$15au (Avenhaus et al. 2014) can now be seen to be asymmetric. The gap clearly deviates from an elliptic shape in the southwest.
{The spiral arms in this region seem to cross the inner wall of the outer disk, extending inwards into the gap region.}

\subsection{Inner dust structure and dust within the gap}
\label{secInnerDust}
\begin{figure*}
\centering
\includegraphics[width=0.97\textwidth]{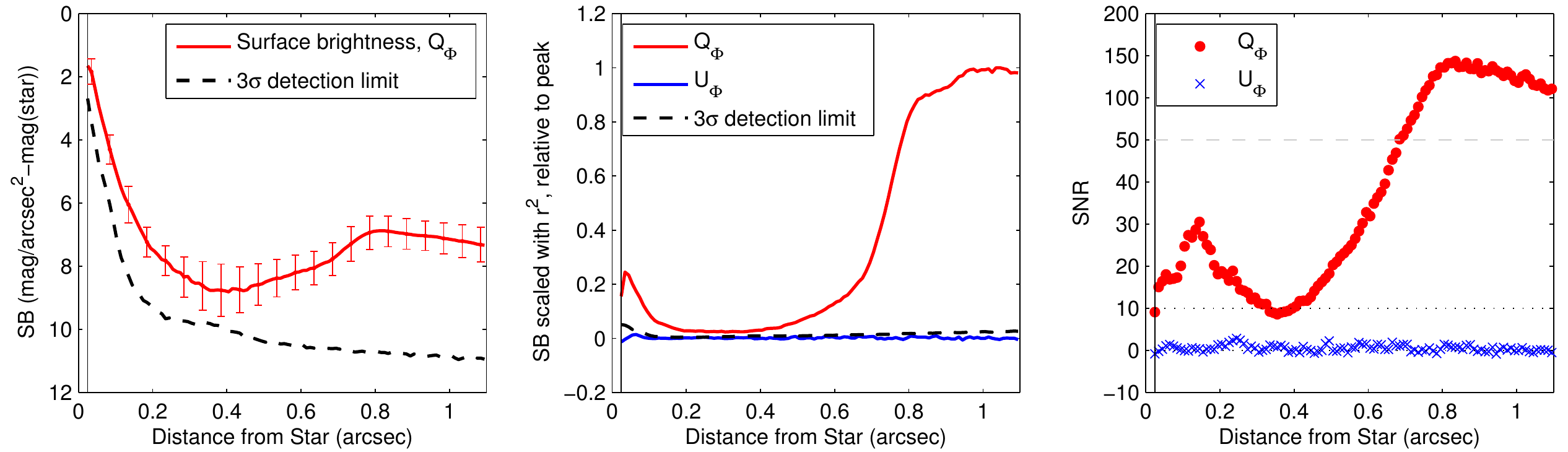}
\vspace{8pt}
\caption{Surface brightness (azimuthal average) and detection strengths of the polarimetric signal. Left: Surface brightness of $Q_{\phi}$ relative to the detection limit. {The error bars show the internal variation along the azimuthal ring, as calculated from the standard deviation within the respective ring. This does not take into account the flux loss due to PSF smearing (see below)}. Middle: $Q_{\phi}$ and $U_{\phi}$ brightnesses after scaling with $r^2$, relative to the peak brightness. While the surface brightness of the inner dust is much higher than that of the outer disk, the scattering strength seems weaker. Right: Detection strength (SNR) determined from comparing to the $U_{\phi}$ data (conservatively assumed to contain random noise). Within the whole range (IWA of 25 mas to 1.1 arcsec displayed here), $Q_{\phi}$ is detected everywhere with at least 8.5$\sigma$ confidence (the dotted line represents 10$\sigma$). Note the change of scale at 50$\sigma$.}
\label{figSurfBrightness}
\end{figure*}

In order to further determine the reliability of the detected signal, we calculate azimuthally-averaged surface brightness profiles for the disk. The results can be seen in Figure \ref{figSurfBrightness}. The polarization signal close to the star is detected at more than 20$\sigma$, with $\sigma$ calculated from the $U_{\phi}$ data as described in \citet{avenhaus2014a}. This does take into account the (systematic) errors close to the star discussed above. We can thus clearly state that the inner dust structure is detected.

\begin{figure}
\centering
\includegraphics[width=0.47\textwidth]{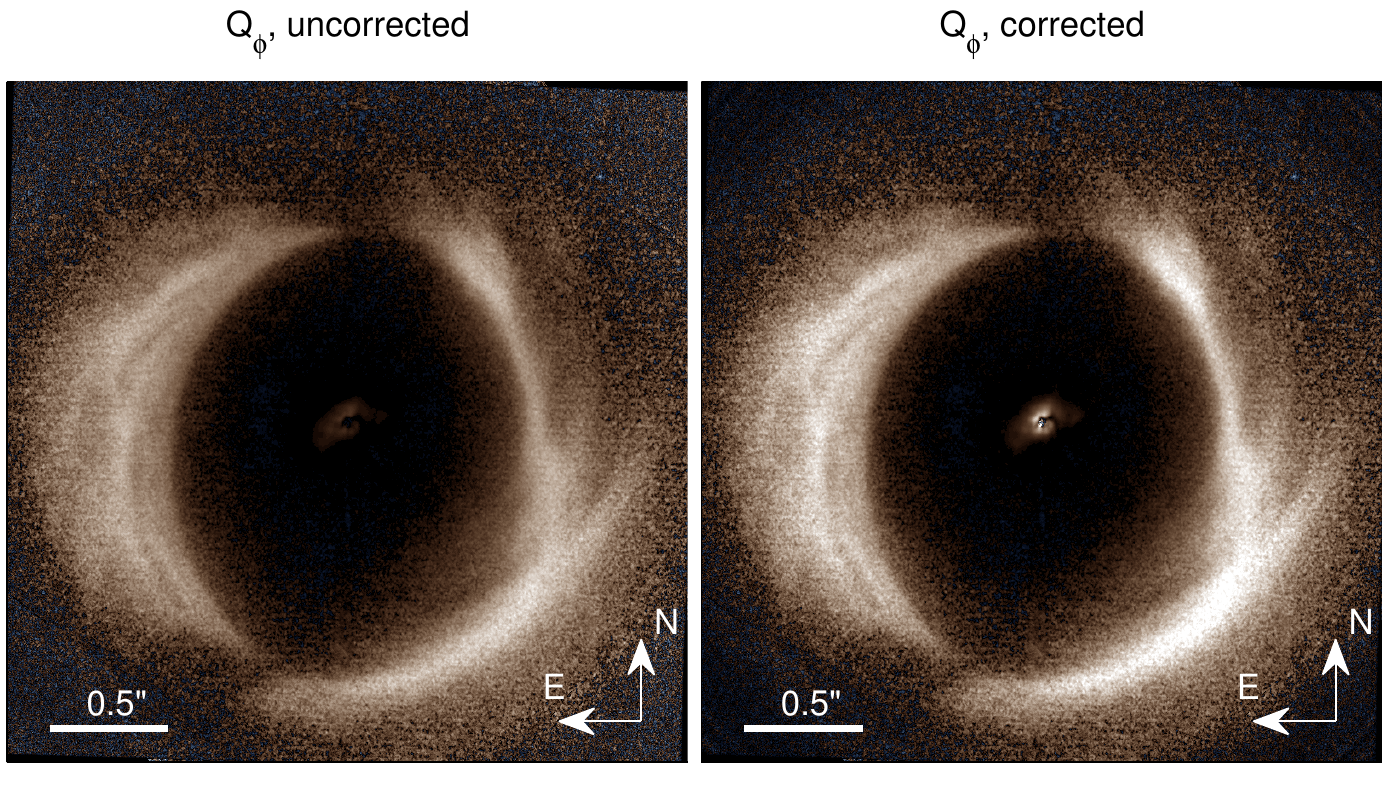}
\vspace{8pt}
\caption{Test of the severity of the signal suppression close to the star (description in main text). After correction, the inner dust structure brightens up significantly, getting close in brightness to the outer disk in this $r^2$-corrected images. The outer disk gets brighter, too, showing that signal suppression due to PSF smearing plays a role even at $>$1$\arcsec$.}
\label{figSupTest}
\end{figure}

\subsubsection{Signal dampening through PSF smearing effects and brightness of the inner disk}

It is worth noting that when corrected for the fall-off of the stellar illumination by multiplying the data with $r^2$, $r$ being the projected distance from the stellar position (right side of Figure \ref{figSurfBrightness}), the dust scattering close to the star seems to be significantly fainter than the outer disk. However, this does not take into account the dampening effect of PSF smearing in PDI. This can reduce the polarimetric flux close to the star when employing the PDI method \citep[see][]{avenhaus2014b}. The magnitude of this effect depends both on the distribution of scattered light itself and on the shape of the PSF. It is weaker for stable, high-Strehl PSFs and further away from the star. The ZIMPOL Very Broad Band filter (590-881nm) is more strongly affected by this problem compared to the near-IR IRDIS filters because of the significantly lower Strehl ratios at this wavelength. The inner dust structure is particularly affected because of its proximity to the star.

In principle, the best way to understand these effects is a forward-modeling of scattered-light images produced with a radiative transfer code, which are then (Stokes Q and U vectors) convolved with the PSF retrieved from the observations. Because developing a radiative transfer model is beyond the scope of this paper, specifically for the complex asymmetric dust structure we observe, we instead use the derived $Q_{\phi}$ image in order to estimate the magnitude of signal suppression within our scattered-light data. The process works as follows: 
\begin{enumerate}
\item Produce an azimuthally averaged image $Q_{\rm \phi, avg}$ of the $Q_{\phi}$ image and smooth it with a small ($\sim$25 mas) Gaussian kernel in order to avoid effects from small-scale structures and noise
\item Split this image into the Stokes Q and U vectors using the inverse of the formulas shown in Section \ref{Observations}
\item Convolve the obtained Stokes vectors with the PSF obtained from the unsaturated science frames
\item Calculate $Q_{\rm \phi, avg, damp}$ from these convolved Stokes vectors
\item Calculate the local damping factor as $F_{\rm damp} = \frac{Q_{\rm \phi, avg}}{Q_{\rm \phi, avg, damp}}$. We then get an approximation of the real (undamped) polarimetric scattered-light signature by multiplying $Q_{\phi}$ with $F_{\rm damp}$.
\end{enumerate}

\begin{figure*}
\centering
\includegraphics[width=0.95\textwidth]{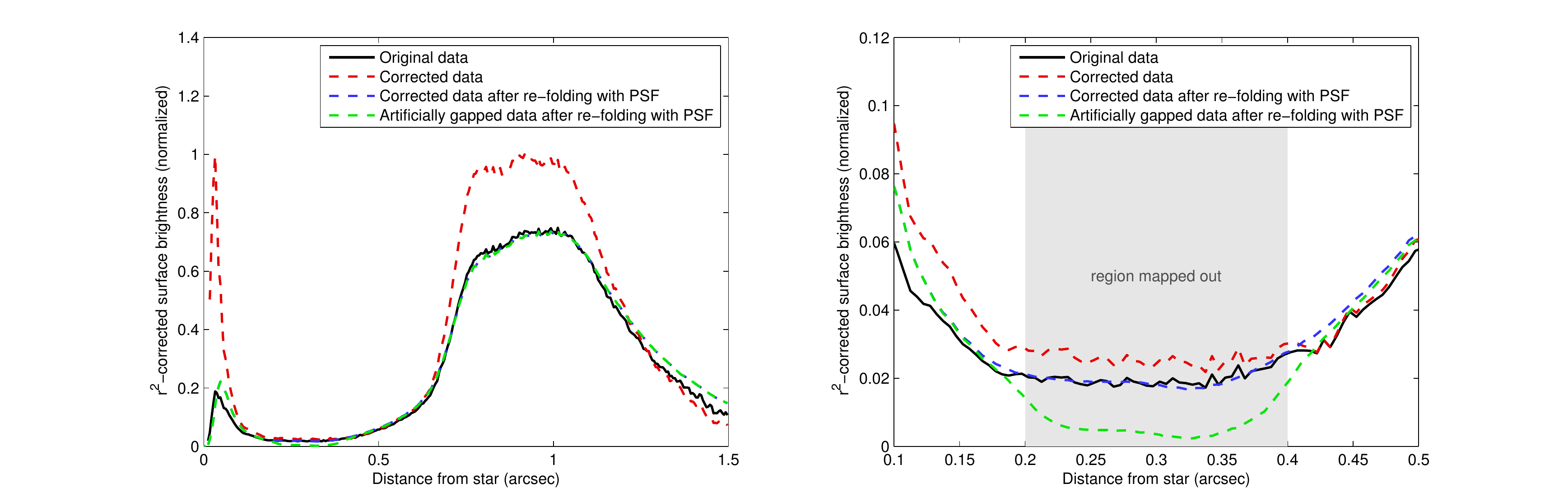}
\vspace{8pt}
\caption{Radial surface brightness of several versions of the $Q_{\phi}$ data compared (all scaled with $r^2$). We compare the original data with the corrected data as well as the corrected data after re-folding with the PSF. Correcting the data shows that the inner disk is indeed much brighter than it seems to be in the original, uncorrected data. Re-folding leads to data that very closely resembles the original data (small differences are due to the additional smoothing from the re-folding with the PSF, resulting in the flux being effectively smoothed twice by the PSF and shifting flux outwards - the inner disk also does not encompass more flux, but only seems to do so due to the $r^2$ scaling, pronouncing the flux that is now further out), showing the validity of our corrections. Additionally, we test with a model where an artificial gap has been introduced between 0.2\arcsec and 0.4\arcsec, to check whether the flux within the gap can be explained by PSF-smearing from the outer or inner disk. The right side shows a zoom into the gap region.}
\label{FigFluxLeakage}
\end{figure*}

Both $Q_{\phi}$ and $Q_{\phi} \cdot F_{\rm damp}$ are displayed alongside each other in Figure \ref{figSupTest}. As can be seen, the inner dust structure brightens up significantly with this processing (factor of $\sim$5). The outer disk brightens as well (showing that PSF smearing has an effect even at $>$1$\arcsec$), but by a smaller factor. We then produce averaged radial surface brightness plots again, which show that the inner disk is indeed as bright as the outer disk when corrected for the drop-off in stellar illumination (see Figure \ref{FigFluxLeakage}, left side).

\begin{figure}
\centering
\includegraphics[width=0.47\textwidth]{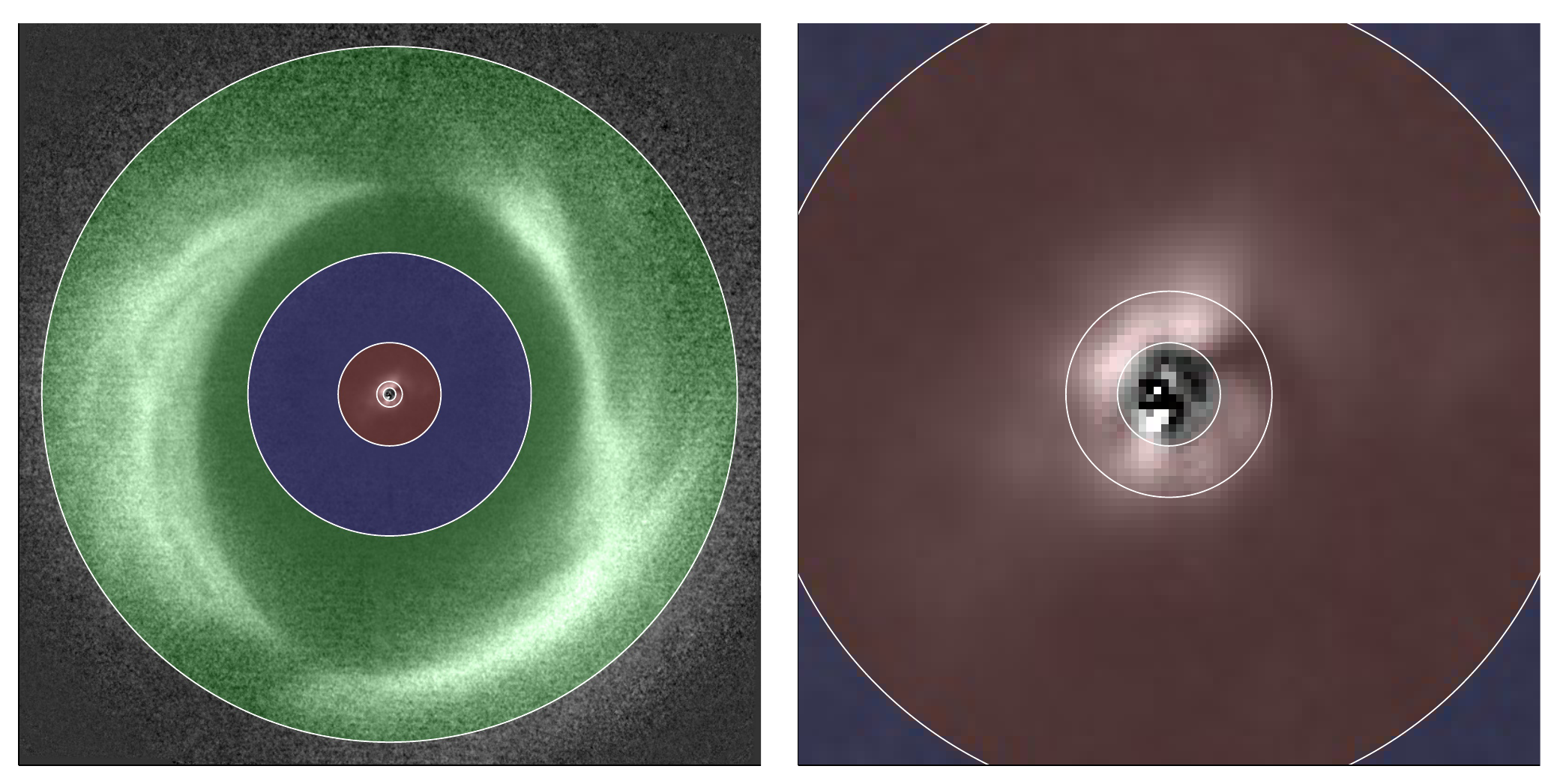}
\vspace{8pt}
\caption{The disk regions we choose to compare. Red: Inner disk, blue: Gap region, green: Outer disk. The right size shows a zoom into the inner region, with a sub-division at 50mas. Below 25mas separation, the data is affected by residuals too strongly to be useful (grey region in the middle).}
\label{figDiskRegions}
\end{figure}

{The inner dust structure is in fact very bright. When compared to the outer disk, it scatters \textit{more} light than the entire outer region of the disk as seen in our images. To show this, we divide our image into three regions: The inner dust structure (0.025\arcsec-0.2\arcsec), the gap region (0.2\arcsec-0.55\arcsec), and the outer disk (0.55\arcsec-1.35\arcsec). Using the corrected $Q_{\phi}$ image and conservative error estimates constructed from the corrected $U_{\phi}$ image, we calculate a polarized flux ratio of $\frac{F_{\rm inner}}{F_{\rm outer}} = 1.43\pm0.36$. Most of this flux is very close to the star, in the region between 25mas and 50mas. If we further divide the inner disk based on this into a region of 25mas-50mas and 50mas-200mas, we get $\frac{F_{\rm 25-50}}{F_{\rm outer}} = 1.04\pm0.23$ and $\frac{F_{\rm 50-200}}{F_{\rm outer}} = 0.37\pm0.07$. The gap is dark compared to the outer disk, with $\frac{F_{\rm gap}}{F_{\rm outer}} = 0.062\pm0.007$. Figure \ref{figDiskRegions} shows the chosen regions of the disk for reference.}

{Comparing the scattered light within 1.35$\arcsec$ to the total flux (star and scattered light, both polarized and unpolarized) in the same region, we get a ratio of $F_{\rm pol, scattered}/F_{\rm total} \approx 0.83\%$. This ratio is affected by both the albedo of the grains and the polarization fraction. The scattered, unpolarized light is also seen in the pure intensity image. 

{Lacking a suitable reference PSF for subtracting the stellar halo, we can not repeat the analysis performed in \citet{canovas2013} and \citet{avenhaus2014a}. However, a rough estimate based on subtracting scaled versions of the $Q_{\phi}$ image from the intensity image points to the polarization fraction showing qualitatively the same behavior as noted in those works - the polarization fraction is higher on the eastern and lower on the western side.}}

\subsubsection{Dust within the gap}

Besides the structure within the innermost $\sim$200 mas, we also detect {a signal in $Q_{\phi}$ at $>$8$\sigma$ throughout the gap, consistent with previous findings in \citet{avenhaus2014a}, where the signal-to-noise ratio was however much weaker ($<$3$\sigma$). There are two possible sources of this signal: It can either arise from light from the outer or inner disk that has been smeared into the gap region by the PSF, or it is scattered light from within the gap region.}

{In order to distinguish between these two possibilities}
, we artificially map out a gap between 0.2\arcsec and 0.4\arcsec in our corrected data, setting this region to zero, and fold it (Stokes Q and U individually) with the PSF. The result is displayed in Figure \ref{FigFluxLeakage} alongside the other surface brightness plots. While some seeping in occurs, this could account for a signal on the order of $\sim$0.5$\%$ of the peak of the outer disk (in $r^2$-scaling), but not for the observed signal of $\sim$2$\%$ of the outer disk peak. {We thus conclude that most of the observed signal ($\sim75\%$) must come from dust scattering in this region, and only a minor fraction ($\sim25\%$) of it results from PSF smearing.}

{This means that dust scattering within the gap region is detected at a significant level of $\sim6\sigma$ after taking into account possible smearing of signal from the in- or outside of the gap. The gap is not completely devoid of dust, as already hinted at in \citet{avenhaus2014a}. The dust scattering is weak (the signal is weaker than the signal from the outer / inner disk by a factor of $\sim$50-70} when corrected for the r$^2$ drop-off in illumination), meaning that the dust is either shadowed or optically very thin.

\begin{figure*}
\centering
\includegraphics[width=1\textwidth]{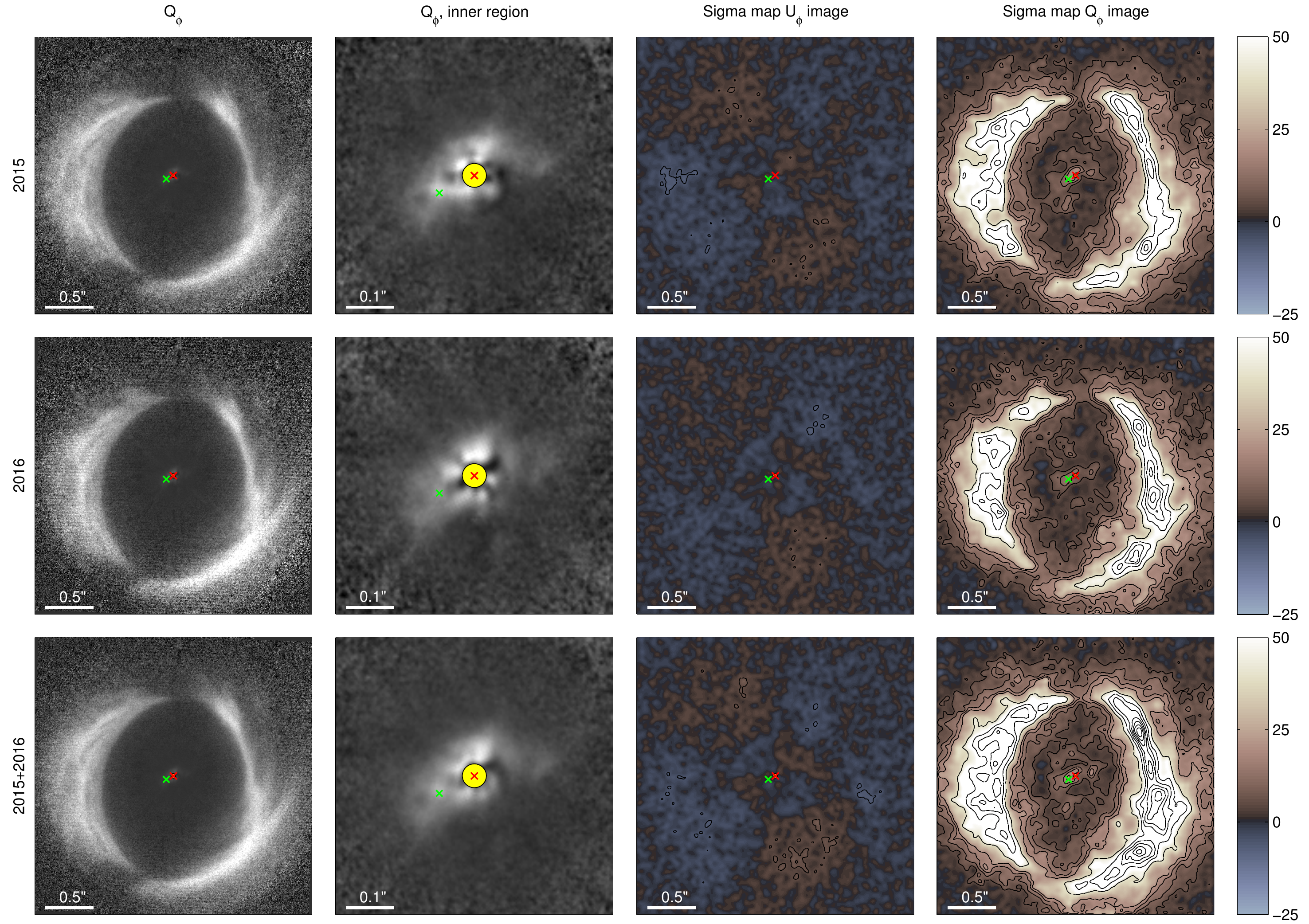}
\vspace{8pt}
\caption{Comparison between the 2015 and 2016 data and approximate detection confidence levels across the image (generation described in main text). Contours are displayed at -5, 5, 10, 15, 20, 40, 60, ... $\sigma$. It can be seen that these sigma maps are not perfect (residual structure and 5$\sigma$ detections in U$_\phi$ images), but overall represent the detection confidence rather well. The small red cross marks the position of the primary in all images, the green cross the position of the secondary on May 12, 2014, \citep{lacour2016}. North is up and East is to the left. The yellow circle marks the region within 25mas, which is dominated by noise.}
\label{figSigmaMaps}
\end{figure*}

In an attempt to better understand the dust distribution within the gap, we create $\sigma$ confidence maps by comparing the signal in the Qr image with the noise in the Ur image \citep[c.f.][]{benisty2015}. These images are generated as follows: 1) Smooth the image with a Gaussian kernel of 30 mas width (approx. the size of the beam) in order to get rid of high spatial frequency noise, 2) Calculate the variance in the (smoothed) Ur image over an area five times as wide as the smoothing kernel, 3) divide the signal in Qr by the square root of the obtained variance. The results are shown in Figure \ref{figSigmaMaps}. This technique is not perfect, namely because the results depend on the used smoothing kernel for the variance calculation, but allows us to give an estimate of the signal strength within the gap. 

We notice that {there is significant scattering close to the outer edge of the gap}, potentially from gas streaming into the gap and carrying small dust grains with it. {This can also be seen from the spiral-like features at the outer edge of the gap in the southwest,} consistent with the fact that the inner disk would not be able to support the accretion rate of the star for extended periods of time and thus material has to accrete from the outer to the inner disk \citep{verhoeff2011}. This feature could not be clearly seen in previous observations \citep{avenhaus2014a}. The signal of the outer disk is very strong (up to 150$\sigma$). {On the other hand, locating the dust within the gap, while clearly detected in the azimuthal averages, remains fairly inconclusive given the low signal-to-noise ratio and the fact that butterfly patterns can be easily introduced into $Q_{\phi}$ and $U_{\phi}$ images if the correction for instrumental polarization is not perfect (these do not affect azimuthal averages, as the positive and negative butterfly wings cancel out). What can be seen, though, is that neither in the 2015 nor in the 2016 or combined data, any shadow can be seen in the direction of the outer disk shadows.}

\subsection{Comparison of 2015 and 2016 epochs}
\label{sec:2015vs2016}

The 2015 and 2016 epochs have similar SNR across the disk and thus can be compared well. The appearance of the disk is very similar, as is to be expected, and combining the data increases the SNR of both the inner dust structure and outer disk (see Figure \ref{figSigmaMaps}). However, there are minor differences worth pointing out.

First, the inner dust structure, while appearing elongated in the SE-NW direction in both epochs, appears broader in the 2016 images in the N-S direction. This can be seen in both the SE and NW extension (Figure \ref{figSigmaMaps}) and is independent of reduction parameters. Second, the structure of the gap in the $\sigma$ maps is different. While there seems to be a bridge-like structure extending from the SE to the NW in the 2015 dataset, the opposite is true for the 2016 epoch - the bridge extends from the NE to the SW. {However, as pointed out in the previous section, butterfly patterns can easily accidentally be introduced into these images. The exact structure depends on the reduction parameters, specifically the inner and outer radius used for the instrumental polarization correction. While for any specific set of parameters used, the dust appears differently in the two epochs, we are not convinced that these differences are significant.}

\section{Discussion}\label{sec:HD100546discussion}

We have presented the first unambiguous detection of scattered light off circumprimary dust within 30au of HD142527. That dust must exist at these separations was already known \citep{verhoeff2011}, but it had so far escaped direct detection in scattered light because of its proximity to the star, the required contrast performance, and the fact that the PSF smearing effect strongly suppresses the polarization signal at these separations. Our study is also an attempt at localizing scattered light within the large gap.

\subsection{Inner disk or dust structure and relation to shadowing}

\citet{marino2015} linked the two prominent local minima in the north and south of the outer disk to a shadowing from an highly inclined inner disk. Within the context of our observations, this presents two challenges: 1) the inner dust structure we observe does not resemble a disk that could conceivably be responsible for that shadowing; 2) no shadow is observed within the gap. 

{\citet{verhoeff2011} find that to reproduce their data, they need to invoke not only an inner disk (which is not inclined in their model due to the knowledge at the time), but also a dust halo close to the star. \citet{min2016} show that in order to fit the SED using a two-component (inclined inner and outer disk) model without a halo, they need to extend the scale height of the inner disk beyond the expected hydrostatic equilibrium in order to fit the near-IR flux. They do not discuss, but also not exclude the possibility of a halo which would enhance the near-IR flux. We furthermore know from \citet{lacour2016} that the secondary M-dwarf HD142527B is likely to be in an orbit co-planar with the suggested, highly inclined inner disk. This means that what we observe could in fact be both the inner disk or an extended, not necessarily spherical halo. We note at this point that \citet{verhoeff2011} pointed out that a halo could be replenished by a highly excited debris disk. The stellar companion of HD142527 was not known at the time, but would excite any debris material in the inner disk. A dust halo producing grey extinction also helps to explain the unusually high infrared flux ratio of F$_{\rm IR}$/F$_{\rm star}$ = 0.92.}

{In this interpretation, the inner disk would then still be within or very close to our IWA, and the east-west extended structure seen would not be part of the inner disk, but part of the (dynamically excited) halo. }
This does not explain the "missing" shadow in the gap, but we have to remember that the scattering signal we observe, while a clear detection when azimuthally averaged, is locally still of very low SNR.

{While the dust scattering seems to be faint at first sight, our ad-hoc correction for PSF smearing effects reveals that the polarimetric flux from the inner disk region is in fact greater than that of the outer disk. Inclination effects could play a role here, and we do not know the line-of-sight position of the inner dust, but we do know that the scattering angles in the outer disk are close to 90$^{\circ}$, the ideal scattering angle for strong polarization, due to the low inclination of the disk. Dust grain properties (e.g. a higher albedo) could also play a role, but to first order we have to assume that as much or more light is re-processed in the inner dust structure than is in the outer disk. Thus, the inner dust structure must be close to optically thick - at least very close to the star, where most of the polarized flux stems from. }

\begin{figure*}
\centering
\includegraphics[width=.9\textwidth]{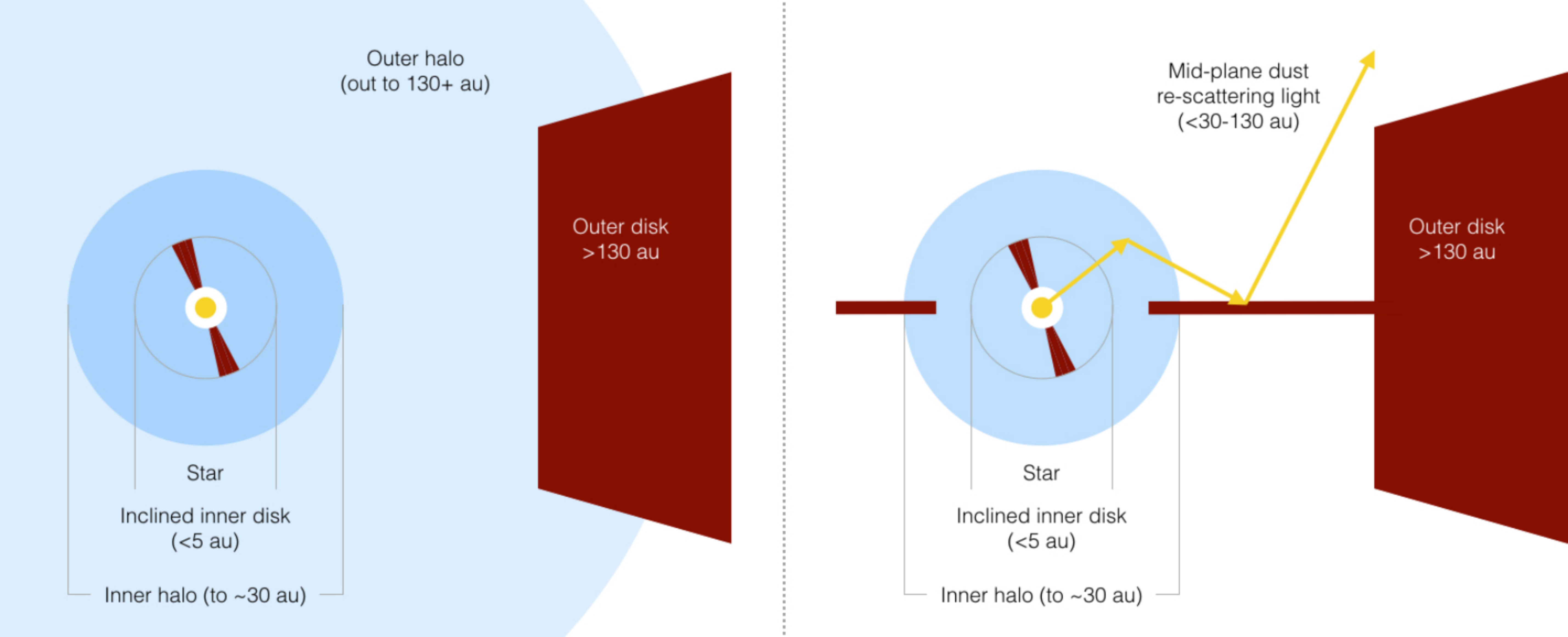}
\vspace{8pt}
\caption{{Two possible explanations for our observations, inspired by \citet{verhoeff2011}. Left: Small inclined inner disk combined with a dense inner halo (to explain the dust scattering seen close to the star) and a larger, faint halo to explain the scattered-light signal in the gap region. Right: Alternative explanation where re-scattering from dust along the mid-plane explains the scattered-light signal in the gap region. These explanations are not meant to represent an exhaustive search for possible explanations for our observations. Representation not to scale.}}
\label{figCartoons}
\end{figure*}

{Combining these findings with our results, we deem the following scenario conceivable: 1) A highly inclined inner disk close to the star, potentially at least partly responsible for the high polarimetric flux very close to the star (25-50mas); 2) A dust halo which is responsible for the weaker polarimetric signal, mostly in the $\sim$50-200mas regime and where the unexpected east-west extension is explained by the fact that it is strongly distorted by interactions with the secondary M-dwarf and material accreting from the outer disk; 3) The well-known outer disk which re-processes large portions of light beyond 100au. {The proposed size of the inner disk is consistent with recent sub-mm observations, in which an inclined disk with a radius of only 2-3 au is sufficient to explain the sub-mm measurements \citep{boehler2017}}.

This scenario, however, does not explain the scattered light seen within the gap nor does it explain the lack of shadows therein. We thus could imagine as component 4) a larger-scale dust halo with dust not only located in the plane of the outer disk, but also high above the mid-plane which would dilute the shadows from the inner disk and thus be compatible with polarized flux from the entire gap region. {The scenario would be similar to the scenario described by \citep{verhoeff2011}, but with a smaller inner disk and an additional large, diluted halo of unexplained origin (it could, for example, stem from radiation pressure blow-out of small dust grains from the inner halo).}

We caution though that our data within the gap is of low signal-to-noise ratio and potentially affected by systematic effects from the unstable PSF of SPHERE in the optical (relatively, compared to the near-IR). Thus, the faint regions in the disk and specifically the gap are difficult to interpret conclusively, and the explanation for the "missing" shadows in the gap might simply be because of too low SNR. {Ignoring the "missing" shadows, another explanation for faint, scattered light within the disk gap would be secondary scattering, i.e. light scattered in the inner region of the disk towards the mid-plane and then re-scattered by dust along the mid-plane within the gap region. However, \citet{boehler2017} find a contrast ratio in the sub-mm between the outer disk peak and the gap region in excess of 1200, meaning that if there is any dust at the mid-plane in the gap region, it must be very thin. We show sketches of the two proposed possible explanations in Figure \ref{figCartoons}.}}

{While this explanation is compatible with our data, as far as we can see without producing a detailed radiative-transfer model, there could be other explanations for this complex circumstellar environment. The exact geometrical shape of the dust we detect remains unknown, also because of possible projection and complicated shadowing effects. It has to tie in, however, with the known kinematic signatures of the gas as studied by ALMA \citep{casassus2015a, perez2015}. To fully understand this, a coupling of a hydrodynamical model which takes into account both the gas and dust dynamics as well as the orbit of HD142527B with a radiative-transfer code (beyond current state-of-the-art capabilities), or at least either adequate hydrodynamical simulations which can handle both gas and dust or radiative-transfer modeling would be necessary.}

\subsection{Dust around HD142527B}

Contrary to the results of \citet{rodigas2014}, we do not see an enhancement of dust close to the location of the M-dwarf companion detected by \citet{biller2012, lacour2016}. Thus, we can not confirm their findings. {We note that the contrast limits we reach are better (\citet{rodigas2014} did not detect the faint dust scattering we report here) and that quick tests with inserting their detection into our images show that we should have detected such a bright point-like source. A disappearance of a circumbinary disk on such short timescales is unlikely. We therefore can not confirm circumsecondary dust scattering, even though a circumsecondary disk is also needed to fit the SED of the secondary \citep{lacour2016}. As a side note, we would like to mention that a detectable signal would stem from light from the primary scattered off the disk of the secondary. The secondary itself is too faint compared to the primary and the butterfly pattern it would produce in Stokes Q and U would be so small that it would be washed out by convolution with the PSF.}

{It is worth mentioning that \citet{boehler2017} detect an additional point-source of sub-mm emission within the gap, towards the north of the star. A possible explanation for this is dust surrounding a third object within the system, however we do not detect any additional scattered light in this region.}

\subsection{Dust entering the gap from the outer disk}

Besides being able to show unambiguously that scattered light exists within the gap of HD142527, which could also stem from a larger halo as described above, we are also able to tentatively localise the polarized flux. The strongest signal is found close to the outer edge of the gap, which could be caused by gas that is dragged into the gap to cross the gap and eventually accrete onto the star (given the accretion rate of HD142527), and dragging micron-sized dust along.

{However, \citet{lacour2016} found that the M-dwarf companion is unlikely to be responsible for truncating the outer disk (due to inclination), and thus also unlikely to be responsible for dragging dust into the gap via companion-disk interactions, despite the line-of-sight proximity of its possible apocenter and the dust signature entering the gap. Unseen planets within the gap would offer a potential explanation.}

\subsection{Differences between 2015 and 2016 epochs}

{The difference in the appearance of the inner dust and the potential difference in the appearance of dust in the gap between the 2015 and 2016 epoch points to a} possible astrophysical difference between the two epochs. A time baseline of 11 months, while short, is still significant compared to the orbital timescale at the location of the inner working angle ($\sim$4 au, $\sim$5 yr). It is very short compared to the orbital timescale at 0.1$\arcsec$ ($\sim$36 yr), though.

It is possible that variations are caused by differential shadowing from dust inside our inner working angle, closer to the star than the dust structure we observe. Dust at that location would evolve on its own, shorter orbital timescale. However, a variable shadowing would not only influence the gap region, but also the outer disk. Variable shadowing on the outer disk is not observed. Thus, future observations will be required to confirm variations of the observed dust structure and understand their origins.

\section{Conclusion}\label{conclusion}

In this paper, we present new observations of the HD142527 disk with ZIMPOL in the VBB filter. The disk is clearly detected, and for the first time, unambiguous detections of dust scattering within the gap and an inner dust structure are presented.

However, our detection of the dust structure close to the star raises several questions. While it is in reasonable agreement with the model of \citet{verhoeff2011}, it is not easy to reconcile with the findings of \citet{marino2015} and \citet{perez2015}, {which predict an inclined inner disk, which we do not see with the right inclination. It is unclear whether the dust we see resembles a disk, and if it does, the inclination and position angle are such that it is unlikely to be responsible for shadowing the two prominent local minima in the outer disk. A model consisting of both an inner disk and a dust halo, which are overlaid in our images and cannot be properly separated due to limitations in signal-to-noise ratio, resolution and inner working angle, is conceivable and consistent with our data, but we have no proof for such an arrangement of the dust.} 

The detection of the dust within the gap is not easily located, even though it is clearly detected in the azimuthally averaged maps. There is no obvious shadow towards the shadows in the outer disk. {This could be explained by a larger-scale dust halo outside the disk mid-plane, but it is also possible that our signal-to-noise ratio is simply too weak to detect the shadows.} There also seems to be some difference between our two epochs, but this detection at this point is tentative and will require further investigation in the future.

{HD142527 is a system in dynamical evolution displaying complex interplay of its misaligned components - the outer disk, the gap region, an inclined inner disk, a halo, the secondary M-dwarf, and potentially more that we have not discovered yet, for example planets responsible for carving out the large gap. The dust and gas components of the disk are clearly not azimuthally symmetric. Understanding the individual components and their interplay remains a challenging task, but due to the proximity and brightness of the disk, HD142527 remains an important transition disk test case.}

\acknowledgments
HA acknowledges support from the Millennium Science Initiative (Chilean Ministry of Economy) through grant RC130007 and further financial support by FONDECYT, grant 3150643. JH is supported by ANR grant ANR-14-CE33-0018 (GIPSE) and FM by grant ANR-16-CE31-0013 (Planet-Forming-Disks).
	
SPHERE is an instrument designed and built by a consortium consisting
of IPAG (Grenoble, France), MPIA (Heidelberg, Germany), LAM (Marseille,
France), LESIA (Paris, France), Laboratoire Lagrange (Nice, France), 
INAF Ð Osservatorio di Padova (Italy), Observatoire de Gen`eve 
(Switzerland), ETH Zurich (Switzerland), NOVA (Netherlands), ONERA (France) 
and ASTRON (Netherlands), in collaboration with ESO. 
SPHERE was funded by ESO, with additional contributions from CNRS (France), 
MPIA (Germany), INAF (Italy), FINES (Switzerland) and NOVA (Netherlands). 
SPHERE also received funding from the European Commission Sixth and 
Seventh Framework Programmes as part of the Optical Infrared 
Coordination Network for Astronomy (OPTICON)
under grant number RII3-Ct-2004-001566 for FP6 (2004Ð2008), grant number
226604 for FP7 (2009Ð2012) and grant number 312430 for FP7 (2013Ð2016).

Part of this work has been carried out within the framework of the National Centre for Competence in Research PlanetS supported by the Swiss National Science Foundation. H.A., S.P.Q., and H.M.S. acknowledge the financial support of the SNSF. 

This research has made use of the SIMBAD database, operated at CDS, Strasbourg, France. We thank the staff at VLT for their excellent support during the observations.  

This work has made use of data from the European Space Agency (ESA)
mission {\it Gaia} (\url{http://www.cosmos.esa.int/gaia}), processed by
the {\it Gaia} Data Processing and Analysis Consortium (DPAC,
\url{http://www.cosmos.esa.int/web/gaia/dpac/consortium}). Funding
for the DPAC has been provided by national institutions, in particular
the institutions participating in the {\it Gaia} Multilateral Agreement.

{\it Facilities:} \facility{VLT:Melipal (SPHERE)}

\bibliographystyle{apj.bst}
\bibliography{HD142527_2016.bib}

\end{document}